\documentclass[conference]{IEEEtran}
\usepackage{amsmath}
\usepackage{epsfig,subfigure,float,ad}

\begin{document}

\title{Collective Adaptive Systems: \\Challenges Beyond Evolvability}

\author{$^1$Serge Kernbach, $^2$Thomas Schmickl, $^3$Jon Timmis\\[1mm]
\small $^1$Institute of Parallel and Distributed Systems, University of Stuttgart, Germany, \emph{serge.kernbach@ipvs.uni-stuttgart.de} \\
\small $^2$Artificial Life Lab of the Department for Zoology, University of Graz, Austria, \emph{thomas.schmickl@uni-graz.at} \\
\small $^3$Department of Electronics and Department of Computer Science, University of York, UK, \emph{jtimmis@cs.york.ac.uk}
}
\date{}
\maketitle

\begin{abstract}
\footnote{Appeared in the workshop ``Fundamentals of Collective Adaptive Systems'', European Commission, 3-4 November, 2009, Brussels.}This position paper overviews several challenges of collective adaptive systems, which are beyond the research objectives of current top-projects in ICT, and especially in FET, initiatives. The attention is paid not only to challenges and new research topics, but also to their impact and potential breakthroughs in information and communication technologies.
\end{abstract}

\section{Introduction}

Collective systems are one of the most largest classes in nature \cite{Bonabeau99} and technics \cite{Kernbach08}: insect colonies, bird flocks, fish shoals, animal herds, human crowd, cars on streets, computers in internet, cellular phones, molecules in bio-syntectic systems and many other examples. All these domain are "collective systems": social, networked, swarm, collaborative, colloidal, nano and others, however all of them indicate the same essential property: elements provide "more" functionality when they are causally coupled. The value of this "more" depends on technology: many communities world-wide, and especially in Europe, are working on various aspects of collective systems. These communities includes adaptive and bio-inspired systems, evolvable and reconfigurable hardware, biological and bio-syntectic systems, software-intensive and distributed systems as well as various branches of networks and network-based approaches. Joining effort of several such communities, as e.g. collective and adaptive, allows formulating common problems, solutions and challenges as well as essentially increase impact in the field of information and communication technologies.

Currently, technological progress enables a new kind of real-world systems: self-replicating (like bacteria)~\cite{Sylvain09}, huge number (nano- and micro- areas)~\cite{Balzani03}, with a high developmental plasticity (bio-chemical and micro-modular systems) ~\cite{Jaramillo09}, with many self- (healing, maintaining, programming, developing) and so on. Development in networked and software-intensive systems~\cite{Wirsing08}, like mobile telephony, demonstrated a great achievement in terms of flexibility and scalability. One of fundamental research questions in these systems is related to capabilities of their adaptability and self-determining behavior in real hard-to-work environments, like underwater, in space or in hazardous situations. These environments are of huge practical relevance and possess essential market potential, especially for oil-free economics and tomorrow's green technologies.

\begin{figure}[htp]
\begin{center}
{\epsfig{file=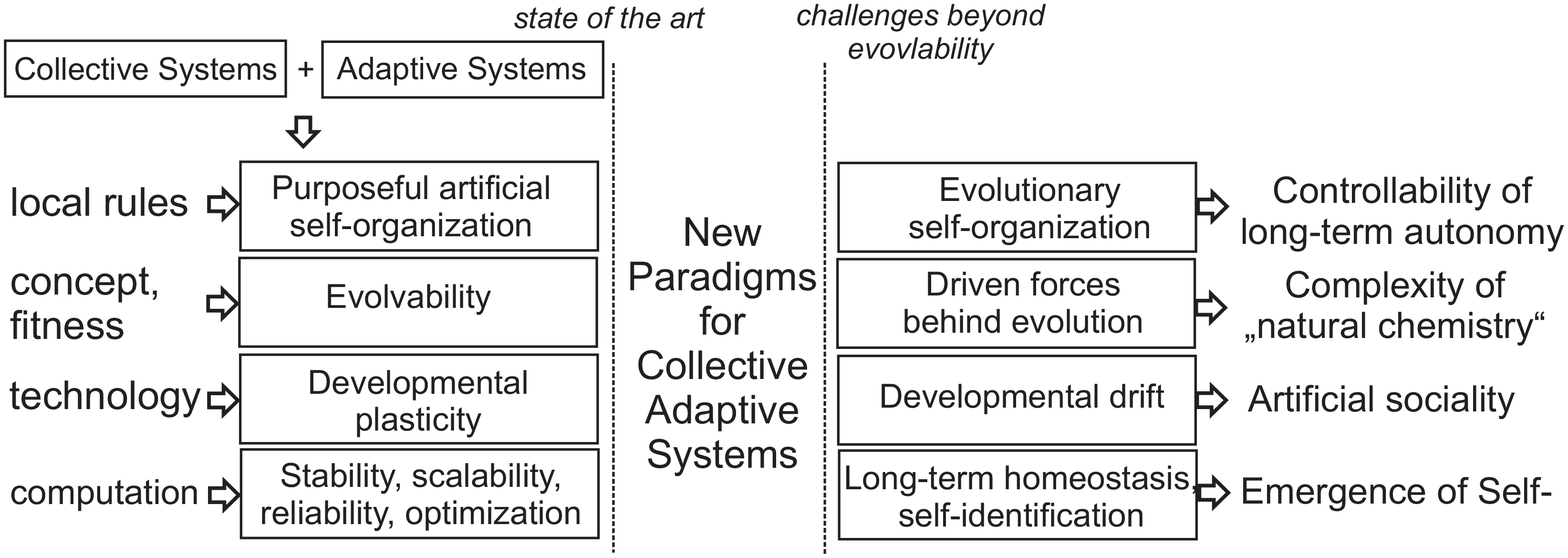,width=.49\textwidth}}
\caption{\em \small Sketch of several challenges beyond the state-of-the-art of current top research projects around collective adaptive systems. \label{fig:fig1}}
\end{center}
\end{figure}

State of the art in the research of collective adaptive systems~\cite{Kernbach10} is very large: it includes bio-inspired and self-organizing branches, evolutionary and adaptive-control strategies, different software and hardware approaches. Approximating a research progress in different top-projects in EU, the point of adaptive systems is focuses on making collective systems cognitive, cooperative, evolvable (both fitness- and concept-driven), self-organizing, with a high developmental plasticity as well as exploring their emerging properties. However, expanding this research mainstream further, after today's start-of-the-art, we are basically missing a few principal cross-domain elements, which are not related to a specific technology, see Fig.~\ref{fig:fig1}.
 The main questions are:
\begin{itemize}
  \item What is beyond adaptability, evolve-ability and emergence of behavior?
  \item What are the driving forces of long-term developmental processes?
  \item Are long-term developmental processes still controllable? Is evolutionary self-organization still controllable?
  \item Is there any developmental drift due to emergence of artificial sociality and self-recognition?
  \item Are there artificial structural elements, which are "absolutely plastic" in the developmental sense, such as biological amino acids?
  \item Is a ``natural chemistry'' (=high complexity of evolutionary processes) important for adaptability and self-development?
  \item Is there an ``artificial chemistry'' that has the ability to to adapt software {\em in-situ}?
  \item Does artificial homeostasis and rules of ecological survival lead to self-identification and to emergence of self-?
\end{itemize}
\vspace{1mm}

\textbf{I. Controllability of long-term self-developmental processes.} The issues of a long-term controllability of autonomous artificial systems is extremely important. Artificial adaptive systems with a high degree of plasticity \cite{Levi10} demonstrate a developmental drift. There are many reasons for this, like long-term developmental independency and autonomous behavior, emergence of artificial sociality, mechanisms of evolutionary self-organization (which are also a huge challenge) and so on. Such systems are very flexible and adaptive, but they also massively increase own degrees of freedom. New challenges in this area are related to a long-term controllability and predictability of "self-", principles of making plastic purposeful systems, predictability of a structural development and goal-oriented self-developing self-organization. These challenges have a great impact on a human community in general (the "terminator" scenario) as well as in different areas of embodied evolution, like synthetic biology or evolvable/reconfigerable systems and networks.
\vspace{1mm}

\textbf{II. Complexity of "natural chemistry".} Coupled with natural chemistry is the development of novel artificial chemistries that have the ability to re-write maybe the operating system, or control system in which it is embodied. In biological systems, the chemical machinery of an organism is data and processor simultaneously, thus providing very complex interaction network but also powerful computational computation. The understanding of such networks is very important as well as the generation of building rules, that allow to build such systems in an engineering way. System dynamics and evolutionary approaches from the field of Artificial Life could to develop such systems more easily. Basically, developing such chemistries is non-trivial: many  such chemistries at the moment suffer from scaling issues, syntactic and semantic problems and general flexibility.  The inclusion of such a chemistry raises its own problems, and indeed would affect to the evolvability and stability of the system, as what is being evolved is also in constant flux.\vspace{1mm}

\textbf{III. Artificial sociality.}
Not only chemical networks become complex quite quickly. Also interaction networks
which arise in social systems can easily get so complex that the cause-and-effect
chains are hidden by the overwhelming network of side-effects and indirect causations. Thus studying such systems, and in parallel, developing the tools needed to \emph{understand} these systems is an important goal. On the one hand, modern technical networks are still not "autonomous" and "scalable" enough to be satisfying, thus investigating comparably complex interaction networks in nature (e.g., social insect colonies) will provide new mechanisms and novel insights, that will help to understand the emerging complexity in modern-world systems. In parallel, creating such systems from scratch (artificial evolution) or form "building blocks" of complexity is also a very promising approach, as long as the products of these "functionality generators" are really investigated and analyzed. Without understanding the "why?" and the "how?" in the evolutionary pathway, these approaches are providing just snapshots and no generalizable insights.
\vspace{1mm}

\textbf{IV. Emergence and controllability of Self-.} Different computational processes, leading to a global optimization, scalability and reliability of collective systems, create a homeostatic regulation. Homeostasis, as well as artificial hormonal regulation, are important and challenging mechanisms in collective adaptive systems. Moreover, conditions of ecological surviving, imposed on such systems, lead to a discrimination between "self" and "non-self" as well as to emergence of different self-phenomena (denoted as "self-"):  self-replication, self-development, self-recovering and other self-. There are several great challenges, like understanding these mechanisms or long-term predictability (see above) which have a large impact on the areas of artificial intelligence and intelligent systems as well as create a new paradigm for adaptive and self-developmental systems.  An additional challenge is to be able to ``engineer emergence'' \cite{polack06}. We envisage systems that are highly evolvable, will adapt themselves over long periods of time, and present emergent properties: todays engineering approaches simply can not address such a challenge.  Ways of controlling emergence in systems, such that we know at least what they {\bf wont} do is essential to constructing systems that might be used in a daily and normal environment.
\vspace{1mm}

The mentioned challenges are only a few very important ones in the area of collective adaptive systems. They are related to a  large community in Europe and have a great impact in tomorrow's ICT field.

\small

\end{document}